\documentclass[5p]{elsarticle}
\journal{Physics Letters B}
\date{}
\usepackage{epsfig}
\usepackage{multirow}
\usepackage{bm}
\usepackage{latexsym}
\usepackage{url}
\usepackage{balance}
\usepackage[T1]{fontenc}
\usepackage{lmodern}
\usepackage{color}
\usepackage{amsfonts,amssymb,amsmath}
\usepackage{graphicx,epsfig}
\usepackage{psfrag}
\usepackage{float}
\usepackage{subfigure}
\usepackage{subcaption}
\usepackage[citecolor = ProcessBlue]{hyperref}
\hypersetup{colorlinks=true}
\usepackage{mathtools}
\usepackage{enumitem}
\usepackage{float}
\usepackage{mathrsfs}
\usepackage[dvipsnames]{xcolor}
\usepackage[utf8]{inputenc}
\usepackage{comment}
\biboptions{sort&compress} 
\allowdisplaybreaks
\definecolor{darkblue}{rgb}{0.0, 0.0, 0.62}
\definecolor{deepmagenta}{rgb}{0.7, 0.01, 0.7}
\definecolor{darkred}{rgb}{0.55, 0.0, 0.0}
\begin{document}
\title{Birkhoff's Theorem and Uniqueness: A Peek Beyond General Relativity}
\author[1]{Rajes Ghosh}
\ead{rajes.ghosh@icts.res.in}
\affiliation[1]{organization={International Centre for Theoretical Sciences, Tata Institute of Fundamental Research},addressline={Bangalore 560089, India}}%
\author[1,2]{Akash K Mishra}
\ead{akash.mishra@saha.ac.in}
\affiliation[2]{organization={Theory Division, Saha Institute of Nuclear Physics},addressline={1/AF Bidhan Nagar, Kolkata 700064, India}}%
\author[3]{Avijit Chowdhury}
\ead{avijit.chowdhury@rnd.iitg.ac.in}
\affiliation[3]{organization={Indian Institute of Technology Guwahati},addressline={Assam 781039, India}}%
\begin{abstract}
In General Relativity, Birkhoff's theorem asserts that any spherically symmetric vacuum solution must be static and asymptotically flat. In this paper, we study the validity of Birkhoff's theorem for a broad class of modified gravity theories in four spacetime dimensions, including quadratic and higher-order gravity models. We demonstrate that the Schwarzschild spacetime remains the unique Einstein branch solution outside any spherically symmetric configuration of these theories. Consequently, unlike black holes, the breakdown of junction conditions at the surface of the star further implies that the actual spacetime metric outside a horizonless star in these modified theories cannot simultaneously be spherically symmetric and remain within the Einstein branch. This insight offers a unique observational probe for theories beyond General Relativity.
\end{abstract}
\maketitle
\sloppy
\section{Introduction}\label{sect:intro}
Even after a century of its inception, Einstein's theory of General Relativity (GR) remains the most successful classical framework for describing gravitational phenomena across a vast range of length scales. The predictions of GR have been extensively validated using both classical tests and modern observations --- spanning over solar system tests to the unprecedented precision offered by cosmological surveys~\cite{Will:2014kxa}. Lately, these tests involve the state-of-the-art gravitational wave detections by the LIGO-Virgo-KAGRA (LVK) collaboration~\cite{Discovery, PEGW150914, TOGGW150914, GWOSC-3,  LIGOScientific:2020tif, Cahillane:2022pqm} and shadow observations by the Event Horizon Telescope (EHT) collaboration~\cite{Akiyama:2019cqa, EHT-1}. In light of these remarkable developments, significant efforts have been put to investigate possible deviations from GR solutions, particularly from the Schwarzschild paradigm, which has been widely used in both classical and modern tests of GR to model spacetime outside spherically symmetric objects when rotational effects are negligible~\cite{Will:2014kxa}. These investigations are critical as they probe the limits of GR and explore the potential for new physics in extreme gravity environments where deviations from standard GR solutions might become detectable.


The wide-ranging applications of the Schwarzschild metric are primarily supported by two important facts. Firstly, understanding a spherically symmetric solution like Schwarzschild provides valuable input for finding more realistic rotating solutions. Indeed, there exist well-defined techniques, such as the Newman-Janis algorithm~\cite{Newman:1965tw}, to construct rotating solutions from spherically symmetric ones. Secondly, one of the most important results in GR, namely Birkhoff's theorem, establishes the uniqueness of Schwarzschild solution outside a spherically symmetric object~\cite{jebsen1921allgemeinen, birkhoff}. In addition to its foundational role, Birkhoff's theorem has profound theoretical and observational consequences. For example, along with the regularity of the spacetime at the centre, it implies that the metric inside an empty spherical cavity must be the Minkowski\footnote{In the case of a star, however, the internal solution can be found by prescribing a matter-energy tensor. For example, one such well-known solution is the Schwarzschild internal solution with static spherically symmetric perfect fluid inside a star~\cite{Hawking:1973uf}.}. Moreover, in GR, Birkhoff's theorem claims the absence of time-dependent vacuum solutions outside a star due to its spherical pulsations. Hence, such pulsations cannot alter the exterior metric from that of Schwarzschild, thereby producing no gravitational radiation.

Over the years, several generalizations of this novel theorem have extended its domain of applicability. The first of such extension can be understood by a geometric reformulation of the aforesaid statement of Birkhoff's theorem. As shown in Refs.~\cite{Hawking:1973uf, Frolov:1998wf}, the validity of Birkhoff's theorem in GR can be traced back to the presence of an  additional Killing vector field in spherical symmetry. This notion extends beyond the usual vacuum form of Birkhoff's theorem stated above. In fact, a generalized Birkhoff's theorem can be proven even in the presence of matter whose stress-energy tensor obeys $T_{AB} = (T^C_C/2)\, \gamma_{AB}$, where the uppercase Latin indices span over the $tr$-sector of the underlying spherical symmetric spacetime~\cite{Frolov:1998wf}. Also, the usual proof of Birkhoff's theorem in GR requires that the metric be at least twice continuously differentiable ($C^2$) for calculating the Ricci tensor. However, the theorem is also applicable under a weaker assumption of continuous and piecewise differentiability of the metric~\cite{Bergmann1965}. The validity of the theorem has also been extended to include electromagnetic fields, for which Reissner-Nordstr\"{o}m is the unique spherically symmetric solution of the Einstein-Maxwell equations~\cite{d1992introducing}. More generalizations of Birkhoff's theorem involve the inclusion of arbitrary time-dependent conformal factors in the metric (i.e., conformal gravity)~\cite{Riegert:1984zz} and cosmological constant~\cite{Schleich:2009uj}; as well as extensions to lower and higher spacetime dimensions~\cite{Kiem:1993zd, Schmidt:1997mq, Bronnikov:1994ja, Ayon-Beato:2004ehj, Keresztes:2007vv}, and mild perturbations away from exact spherical symmetry and vacuum~\cite{Goswami:2011ft, Goswami:2012jf}.

Since Birkhoff's theorem is a direct consequence of GR field equations,  its applicability in the modified gravity landscape is not generally guaranteed. However, the importance of carefully assessing its validity in beyond-GR theories can hardly be exaggerated. In fact, there are strong theoretical and observational reasons to believe that GR might receive higher curvature corrections~\cite{Penrose:1964wq, Donoghue:1994dn, Wald:1997wa, Peebles:2002gy, Bertone:2004pz, Clifton:2011jh, Barack:2018yly}, which becomes particularly important in strong-field regimes~\cite{Berti:2015itd, Berti:2019tcy}. Thus, it is both interesting and important to study the status of various novel results of GR, like Birkhoff's theorem, within the framework of these modified theories. Such studies can provide insights into potential universal features of the beyond-GR landscape. Motivated by these considerations, Birkhoff's theorem has already been extended to Lovelock gravity in $D>4$ dimensions~\cite{Zegers:2005vx, Deser:2003up, Deser:2005gr} (more generally to quasi-topological gravity~\cite{Oliva:2011xu, Cisterna:2017umf, Oliva:2010eb,bueno2024regularblackholespure}), and for Kleinian geometry~\cite{Easson:2023ytf}. Additionally, Birkhoff's theorem has been explored in the context of scalar-tensor theories~\cite{Faraoni:2010rt}, Einstein-\ae ther theory~\cite{Chan:2022mxd} and Horava-Lifshitz gravity~\cite{Devecioglu:2018ote}.
It is also known to hold in $f(R)$ gravity under certain conditions~\cite{Capozziello:2011wg, 2018EPJP..133..376R, Nzioki:2013lca}. However, a comprehensive approach to extend its validity to other modified theories is still lacking. 

In this work, we aim to bridge this crucial gap for the first time by establishing Birkhoff's theorem for a wide class of beyond-GR theories in $4$-dimensions (4D). The Lagrangians of these theories involve both Ricci scalar and Ricci tensor invariants, which will be discussed in more detail in the subsequent sections. Unlike in GR, such modified theories lead to multiple branches of solutions, which may be either analytic or non-analytic functions of higher curvature coupling constants. However, among them, the so-called \textit{Einstein branch} has an analytical structure for all values of the higher curvature coupling constants and, especially in the limit of vanishing coupling constants, this branch is smoothly connected to GR. In other words, all geometrical quantities (such as metric components and curvatures) in this branch can be written as a linear combination of analytical terms in couplings\footnote{Sometimes various physical quantities are expressed in terms of transformed variables using field redefinitions. For example, in theories like $f(R)$-gravity, one can work either in the Jordan frame or in the Einstein frame (not to be confused with the Einstein branch), after doing a conformal transformation of the metric $g_{\mu \nu} \to \tilde{g}_{\mu \nu}$. In such scenarios, the notion of the Einstein branch boils down to an equivalent notion that physical quantities must be analytic in all non-GR variables, if they themselves are analytic in coupling constants. However, as a cautionary note, the ``equivalence'' between the Einstein and Jordan frames is a much-debated topic in literature~\cite{Capozziello:2010sc,Kamenshchik:2016gcy}. Especially in the context of Birkhoff's theorem, which we shall study in the Jordan frame, its equivalent formalism in the Einstein frame is not obvious and lies outside the domain of the present work.}. Furthermore, stationary BHs in the Einstein branch are promising thermodynamic candidates beyond GR, as they have been shown to satisfy the zeroth law~\cite{Ghosh:2020dkk, Bhattacharyya:2022nqa}. These remarkable properties are evident for the astrophysical relevance of the this branch.

These novel features motivate us to explore the extension of Birkhoff's theorem within the Einstein branch of a class of modified theories including quadratic gravity (QG) discussed in the subsequent sections. In these theories, due to the presence of helicity-$0$ modes, the Birkhoff's theorem fails to hold in general (see for e.g. Ref.~\cite{Xavier:2020ulw}). Hence, it is important to investigate for unique solutions of these theories, even with restricted assumptions. In this spirit, we shall show that Schwarzschild solution remains the unique vacuum solution outside spherically symmetric objects in the Einstein branch of these theories. Similar to the case of GR, the staticity and asymptotic flatness of the exterior spacetime follow from the spherical symmetry. 

We must emphasize the non-triviality of our result. First, it is not a-priori guaranteed that the Einstein branch of an alternative theory, like QG, would possess only a single type of spherically symmetric solutions. Therefore, our demonstration of the Schwarzschild metric's uniqueness within this branch is significant. Secondly, note that our assumption of the Einstein branch is not to neglect the effect of the helicity-$0$ mode present in such theories. In fact, its effect is there in all branches, as it is one of the degrees of freedom of the theory. The only speciality of the Einstein branch is that the effect of helicity-$0$ mode in this branch creates a smooth analytic deviation from GR. In all other branches, it produces non-analytic departures. 

We shall also discuss several important implications of this result, which are of utmost observational relevance. For instance, we shall argue that as a direct consequence of this powerful theorem, the actual spacetime metric outside a horizonless star in the theories under consideration cannot be both spherically symmetric and remain within the Einstein branch due to the failure of junction conditions at the surface of the star~\cite{khakshournia2023art}. In other words, unlike GR, the spacetime outside a spherically symmetric star in these theories must deviate from the Einstein branch. This deviation, in turn, provides a novel observational probe into the nature of beyond-GR theories.

Throughout this work, we shall set $G=c=1$. We begin our analysis with QG in Sect.~\ref{sect:QG} as a warm-up,  which we then generalize to more comprehensive beyond-QG theories in Sect.~\ref{sect:beyondQG}. Finally, we conclude in Sect.~\ref{sect:conclusion} by outlining potential future generalizations of our findings.

\section{Birkhoff's theorem in the Einstein Branch of QG}\label{sect:QG}
According to the Gauss-Bonnet theorem, the most general gravity theory in $4$-dimensions including terms up to quadratic order in curvatures is the so-called \textit{quadratic gravity} (QG) with the Lagrangian density~\cite{stelle1977PRD,stelle1978GERG}: $\mathcal{L} = R+\alpha\, R^2+\beta\, R_{\mu \nu}R^{\mu \nu}$. While QG is traditionally considered problematic due to the presence of massive spin-2 ghost modes (suggesting perturbative instability), it can still be treated as a ``healthy'' classical theory of gravity. In fact, some recent works have demonstrated that QG is free from causality issues~\cite{Edelstein:2021jyu}, positive energy theorem holds (implying the nonlinear stability), and possesses well-defined dynamical properties~\cite{Held:2023aap}.

It is crucial to recognize that QG, being a higher-curvature theory, may have multiple branches of solutions~\cite{Holdom:2002xy, Daas:2022iid, Lu:2015psa, Lu:2015cqa, Lu:2015tle, Kokkotas:2017zwt, Pravda:2016fue, Lu:2017kzi, Podolsky:2019gro}(see also~\cite{Held:2022abx}). However, as emphasized in the Introduction section, the remarkable observational consistency of GR dictates that the Einstein branch has strong astrophysical relevance. The solutions in this branch are analytically connected to GR in the limit of $\{\alpha,\, \beta\} \to 0$. Consequently, all geometrical quantities can be expressed as so-called \textit{formal power series} in couplings, i.e., in terms of combinations proportional to $\alpha^{n_1} \beta^{n_2}$, with $n_i$'s being non-negative integers for all values of the couplings. One must note the subtle difference between the above claim and the Taylor expansion around vanishing coupling constants (which is not what we are doing\footnote{Let us emphasize that the defining requirement for physical quantities to lie within the Einstein branch is that they exhibit a smooth, analytic dependence on the coupling constants, particularly in the limit as these couplings approach zero (the GR limit). Consequently, all series expansions in terms of couplings should be treated as formal power series, rather than small-coupling Taylor expansions. However, the requirement of analyticity rules out structures such as $\exp(-M^2/\alpha)$ or $\alpha \log \alpha$, which fall outside the Einstein branch.}).\\

The QG field equations in vacuum are given by $E_{\mu \nu} := G_{\mu \nu}+\alpha\, H_{\mu \nu}+\beta\, I_{\mu \nu} = 0$, where $G_{\mu \nu}$ is the Einstein tensor, and the explicit expressions of $\{ H_{\mu \nu},\, I_{\mu \nu}\}$ can be found in Ref.~\cite{stelle1978GERG}. The most general spherically symmetric solution (characterizing spacetime outside spherical objects) of this theory in $4$-dimensions can be written as follows:
\begin{equation} \label{QGmetric}
    ds^2 = -e^{f(r,t)}\, dt^2 + e^{g(r,t)}\, dr^2 + r^2\, d\Omega^2_{(2)}\, .
\end{equation}
Note that the cross terms such as $g_{t \theta}$, $g_{t \phi}$, $g_{r \theta}$, and $g_{r \phi}$ are zero because of spherical symmetry. Furthermore, the metric can always be diagonalized in the $tr$-sector to eliminate the $g_{tr}$ term. Since we are only interested in solutions in the Einstein branch, one must have for all choices of $(\alpha,\beta)$ that
\begin{align}\label{eq:f-full}
    f(t,r) &= \mathrm{ln}\Big(1-\frac{2M}{r}\Big) + \left(\alpha\, f^{(1)}_{(1,0)}(t,r)+\beta f^{(1)}_{(0,1)} \right) \nonumber\\
    &+\left(\alpha^2\, f^{(2)}_{(2,0)}(t,r)+\beta^2 f^{(2)}_{(0,2)}+\alpha\beta f^{(2)}_{(1,1)} \right)+\cdots~,
\end{align}
\begin{align}\label{eq:g-full}
  g(t,r) &= \mathrm{ln}\Big(1-\frac{2M}{r}\Big) + \left(\alpha\, g^{(1)}_{(1,0)}(t,r)+\beta g^{(1)}_{(0,1)} \right) \nonumber\\
    &+\left(\alpha^2\, g^{(2)}_{(2,0)}(t,r)+\beta^2 g^{(2)}_{(0,2)}+\alpha\beta g^{(2)}_{(1,1)}\right)+ \cdots~.
\end{align}
Here, the subscript represents the power of $\alpha$ and $\beta$ associated with a given term. For brevity, we shall write the above expressions in a compact notation as follows
\begin{align} \label{QGexp}
    f(t,r) = \mathrm{ln}\Big(1-\frac{2M}{r}\Big) + \epsilon\, f^{(1)}(t,r)+\epsilon^2\, f^{(2)}(t,r) + \cdots\, ,\\ 
    g(t,r) = -\mathrm{ln}\Big(1-\frac{2M}{r}\Big) + \epsilon\, g^{(1)}(t,r)+\epsilon^2\, g^{(2)}(t,r) + \cdots\, ~,
\end{align}
where, $\epsilon$ is the book-keeping parameter 
tracking the order of $\{\alpha^n,\, \beta^n\}$ in the above expressions. More explicitly, the term proportional to $\epsilon^n$ contains linear combination of all functions proportional to $\alpha^{n_1}\, \beta^{n_2}$ such that $n_1+n_2 = n$ ($n_i \geq 0$). We have taken the $\epsilon^0$-th order solution to be same as GR as Birkhoff's theorem holds in vacuum GR. Thus, to prove the Birkhoff's theorem in QG, we must first demonstrate that the metric Eq.~\eqref{QGmetric} is actually static. For this purpose, let us now study the properties of various components of the field equations order-by-order in $\epsilon$ and as functions of various metric coefficients in the Einstein branch.

(a) Till $\epsilon^1$ order: In this order, it is trivial to check that $E_{t r} = (\epsilon/r)\, \partial_t g^{(1)}$. Setting this to zero immediately implies that $g^{(1)}(t,r)$ must be independent of $t$. Then, substituting this in the expression of $\partial_t E_{rr} \propto \partial_t \partial_r f^{(1)}(t,r) = 0$ leads to the condition that $f^{(1)}(t,r) = h^{(1)}(t) + j^{(1)}(r)$. Then, one can absorb the $h^{(1)}(t)$ part into a redefined time coordinate up to the first order in $\epsilon$, i.e., $t \to t+\epsilon\, t_{(1)}[h^{(1)}]+\mathcal{O}(\epsilon^2)$. This completes our proof at $\epsilon^1$ order. For more clarity and concreteness, we have included an explicit calculation in terms of $(\alpha, \beta)$ in~\ref{App:A}.

(b) Till $\epsilon^2$ order: Similarly, one can check that $E_{t r} = (\epsilon^2/r)\, \partial_t g^{(2)}(t,r) = 0$. It implies that $g^{(2)}(t,r)$ is independent of $t$. Substituting this back in the expression of $\partial_t E_{rr} \propto \partial_t \partial_r f^{(2)}(t,r) = 0$, we get $f^{(2)}(t,r) = h^{(2)}(t) + j^{(2)}(r)$. Finally, we can absorb $h^{(2)}(t)$ into a redefined time coordinate up to the second order in $\epsilon$, i.e., $t \to t+\epsilon\, t_{(1)}[h^{(1)}]+\epsilon^2\, t_{(2)}[h^{(2)}]+\mathcal{O}(\epsilon^3)$.

In fact, one can inductively show that the above calculation follows at all orders in $\epsilon^n$, for the terms $\alpha^{n_1}\beta^{n_2}$ effectively work as basis in the Einstein branch. This finishes our proof that the metric in Eq.\eqref{QGmetric} must be static.
Then, the question remains that what are these metric coefficients that solve QG field equations? To answer this question, let us again consider the $tt$ and $rr$ components of the QG field equation. Up to quadratic order in $\epsilon$, we get the following expressions:
\begin{align} \label{QGf}
    &g_{tt}(r) \approx -\Big(1-\frac{2M}{r}\Big) - \epsilon\, \frac{g_{c1}-2\, f_{c1}\, M+f_{c1}\, r}{r}\, - \frac{\epsilon^2}{r} \times \nonumber \\
    &\left[f_{c1}\, g_{c1}-M\, (f_{c1}^2+2\, f_{c2})+ f_{c1}^2\, r/2\,+(g_{c2}+f_{c2}\, r)\right]\, , \\ \nonumber\\
    &g_{rr}(r) \approx \Big(1-\frac{2M}{r}\Big)^{-1} - \epsilon\, \frac{g_{c1}\, r}{(r-2M)^2}+ \frac{\epsilon^2}{(r-2M)^3} \times \nonumber \\
    &\kern 10em \left[ r(g_{c1}^2+2\, g_{c2}\, M - g_{c2}\, r) \right]\, .
\end{align}
Here, $g_{c\, n}$ and $f_{c\, n}$ are some constants depending on $\alpha^{n_1} \beta^{n_2}$, such that $n_1+n_2 = n$ ($n_i \geq 1$). One can also check that, till quadratic order, $g_{tt}\, g_{rr} \approx -1 - f_{c1}\, \epsilon - (f_{c1}^2/2+f_{c2})\, \epsilon^2$. Therefore, by a redefinition of the time coordinate (absorbing some constants in $dt^2$), we can make $g_{tt}\, g_{rr}= -1$. In fact, going into this new time coordinate is equivalent to setting $f_{c\, n} = 0$ for all $n \geq 1$. With this rearrangement, the metric components become
\begin{align} \label{QGfg}
    &-g_{tt}(r) = \Big(1-\frac{2M'}{r}\Big) = g_{rr}^{-1}(r)\, ,
\end{align}
where $M' = M -0.5 (g_{c1}\, \epsilon + g_{c2}\, \epsilon^2 + \cdots)$ is the redefined ADM mass. Hence, we have shown that in the Einstein branch, the unique spherically symmetric solution of QG is Schwarzschild. This completes our proof of Birkhoff's theorem in QG. However, we shall see in Sec.~\ref{sect:conclusion} that as a consequence of this result, when complied with the failure of junction conditions at the surface of a horizonless star~\cite{khakshournia2023art}, the actual spacetime metric outside the star in QG cannot be simultaneously spherically symmetric and remain within the Einstein branch.

A few additional comments on this result are in order. It is well known that Schwarzschild metric is a solution of QG. However, what we have demonstrated is that Schwarzschild is the unique spherically symmetric vacuum solution (with staticity arising automatically) within the Einstein branch of QG. Notably, there exists a known spherically symmetric, Ricci-flat solution of QG, which is not analytic as it contains $\beta^{-1}$ terms and hence, it is not in the Einstein branch~\cite{Pravda:2016fue, Lu:2017kzi, Podolsky:2019gro}. Furthermore, taking the trace of the QG field equation, we get $2(3\alpha + \beta) \Box R = R$. Here, the combination $2(3\alpha + \beta)$ is the inverse-squared mass of the helicity-$0$ massive scalar (or breathing) mode present in QG. However, since the solution we have found is uniquely Schwarzschild having $R = 0$ in the Einstein branch, the breathing mode does not induce any modifications to the metric outside a spherical star. This is consistent with expectations, as Birkhoff’s theorem implies that spherically symmetric pulsations cannot alter the outside spacetime. 

\section{Extension beyond QG}\label{sect:beyondQG}
To generalize Birkhoff's theorem beyond QG, we revisit our previous analysis from a different perspective. We consider a class of modified theories with vacuum field equations: $G_{\mu \nu} + \epsilon\, K_{\mu \nu} = 0$, with the assumption that $K_{\mu \nu}$ is a functional of Ricci tensor and its covariant derivatives only, such that $K_{\mu \nu}[g^{Sch}_{\mu\nu}] = 0$. However, this apriori does not guarantee that Schwarzschild is the unique spherically symmetric solution of these theories (which encompass both QG and $f(R)$-gravity) and staticity follows from the above field equations and spherical symmetry. However, it is exactly what we show now using a unified technique. 

To establish the uniqueness of the Schwarzschild solution as the spherically symmetric vacuum solution of this class of modified gravity, we again consider the Einstein branch and work with the same metric ansatz as given in Eq.\eqref{QGmetric} along with Eq.\eqref{QGexp}. In the first order in $\epsilon$, we symbolically write the metric as $g_{\mu \nu}(t,r) = g_{\mu \nu}^{Sch}(r)+\epsilon\, g_{\mu \nu}^{(1)}(t,r)$. Substituting it in the field equations and using the fact that $K_{\mu \nu}[g_{\mu \nu}^{Sch}] = 0$, we get $G_{\mu \nu}[g_{\mu \nu}^{Sch}+\epsilon\, g_{\mu \nu}^{(1)}]=\mathcal{O}(\epsilon^2)$ \footnote{This is distinct from the Taylor expansion about $\epsilon = 0$ as explained earlier.}, which is just GR field equations with the deviated metric. However, since Birkhoff's theorem holds in GR, we must have $g_{\mu \nu}(t,r) = g_{\mu \nu}^{Sch}(r)$ till the first order in $\epsilon$, although the ADM mass will receive $\epsilon$-order correction as shown in the previous section. Then, $g_{\mu \nu}(t,r) = g_{\mu \nu}^{Sch}(r) + \epsilon^2\, g_{\mu \nu}^{(2)}(t,r)$ up to second order. The above argument can be easily extended to all orders in $\epsilon$, completing the proof. Therefore, if a modified theory of gravity supports Schwarzschild metric as a vacuum solution, then Birkhoff's theorem holds and Schwarzschild remains the unique spherically symmetric vacuum solution in the Einstein branch (but see the caveat in Sec.~\ref{sect:conclusion} when complied with the junction condition for a star). 

A few important comments are in order. As a special case of the above construction, it is easy to see that GR vacuum shock wave (i.e., boosted BH solution to velocities tending to light's speed producing a singular localized stress-energy tensor) solution is the unique Einstein branch shock in this class of modified gravity, extending Horowitz's result beyond $\sigma$-model~\cite{Horowitz:1989bv}. Moreover, it follows that if $K_{\mu \nu}[g_{\mu \nu}^{Kerr}] = 0$, then Kerr is the unique stationary and asymptotically flat vacuum BH solution in the Einstein branch of this particular class of modified gravity. Note that, unlike the spherically symmetric case, the stationarity and asymptotic flatness of Kerr does not follow from the field equations and need to be assumed.

\section{Conclusion and Discussions}\label{sect:conclusion}
To date, the extension of Birkhoff's theorem beyond GR remains limited. However, the theoretical importance and observational consequences of thoroughly investigating which modified theories comply with such a theorem cannot be overstated. This work aims to bridge this gap for the first time by establishing Birkhoff's theorem for a broad class of beyond-GR theories, including QG. Our proof is valid in the Einstein branches of these theories, which is smoothly connected to GR in the limit of vanishing coupling constants.

Consequently, we demonstrate that the Schwarzschild solution remains the unique vacuum solution outside spherically symmetric objects in the Einstein branch of these theories. As in GR, the staticity and asymptotic flatness of the exterior spacetime follow as a direct consequence of spherical symmetry. This result has some profound implications. For instance, since the Schwarzschild metric has a vanishing Ricci scalar and tensor, it cannot be matched with any valid internal solution in the case of a horizonless object. Hence, the failure of junction conditions, which is now modified for beyond-GR theories~\cite{khakshournia2023art}, suggest that the outside spacetime of a horizonless star cannot both be spherically symmetric and lie simultaneously in the Einstein branch.

Possible future extensions of our work include relaxing the assumption of asymptotic flatness and studying the validity of Birkhoff's theorem for asymptotically de Sitter or anti-de Sitter-type spacetimes. To this end, we note that our argument needs modification for incorporating cases of non-vacuum solutions and those modified theories that imply $K_{\mu \nu}[g_{\mu \nu}^{Sch}] \neq 0$. Moreover, it will be interesting to study this theorem beyond the Einstein branches of various modified gravity theories, particularly those involving non-topological Riemann terms, where our present method does not straightforwardly apply. A thorough investigation of such cases in the context of Birkhoff’s theorem is an important direction, but one that requires a dedicated and extensive study, which we leave for future our future endeavors.

\section*{Acknowledgements}
We are grateful to the anonymous referees for raising several important points during the review. We thank Timothy Clifton for his valuable comments on an earlier version of this draft. We also appreciate the insightful discussions with Sumanta Chakraborty during the project's initial stage. RG extends his gratitude to Sumanta Chakraborty for the invitation to the Indian Association for the Cultivation of Science (IACS) and to all members of the IACS Physics Department for their warm hospitality during his visit. The work of AC is supported by the National Postdoctoral Fellowship of SERB, ANRF, Government of India (File No.: PDF/2023/000550).

\appendix
\section{Explicit calculation up to first order in $\alpha$ and $\beta$}~\label{App:A}
The $tr$-component of the field equation reads
\begin{equation}
    E_{tr}=\frac{\alpha}{r}\, \partial_t g^{(1)}_{(1,0)}(t,r) +\frac{\beta}{r}\, \partial_t g^{(1)}_{(0,1)}(t,r)=0.
\end{equation}
By virtue of Einstein branch, this equation must hold for all value of $\alpha$ and $\beta$. Hence, one must have,
\begin{equation}\label{eq:A-g1dt}
    \partial_t g^{(1)}_{(1,0)}(t,r)=0=\partial_t g^{(1)}_{(0,1)}(t,r)~.
\end{equation}
Taking cognizance of Eq.(\ref{eq:A-g1dt}), we can also write for the $rr$-component of field equation that
\begin{equation}
    \partial_{t}E_{rr}= \frac{1}{r}\partial_t \partial_r \left( \alpha\, f_{(1,0)}^{(1)} (t,r)+ \beta\, f_{(0,1)}^{(1)} (t,r)\right)=0~.
\end{equation}
Again, by virtue of Einstein branch, we obtain
\begin{align}\label{eq:A-f10f01}
 f_{(1,0)}^{(1)}(t,r)=h_{(1,0)}^{(1)}(t)+j_{(1,0)}^{(1)}(r)\\
 f_{(0,1)}^{(1)}(t,r)=h_{(0,1)}^{(1)}(t)+j_{(0,1)}^{(1)}(r)
\end{align}
For brevity, we denote $\alpha \, h_{(1,0)}^{(1)}(t) + \beta\, h_{(0,1)}^{(1)}(t)\equiv \epsilon\, h^{(1)} (t)$. Now, by absorbing $h^{(1)}(t)$ in a rescaled time coordinate proves the Birkhoff's theorem at the first order in $(\alpha, \beta)$. 
\bibliographystyle{elsarticle-num}
\bibliography{ref}
\end{document}